\documentclass[aps,11pt,prd,notitlepage,tightenlines,nofootinbib,showpacs,superscriptaddress,twocolumn]{revtex4-1}
\usepackage{amsmath, amssymb, amsthm}
\usepackage{mathtools}
\usepackage{mathrsfs}
\usepackage{bm}
\usepackage{slashed}   
\DeclareMathAlphabet{\mathpzc}{OT1}{pzc}{m}{it}

\usepackage{graphicx}
\usepackage{multirow}
\usepackage{tikz}
\usepackage{subfigure}
\usepackage{relsize}	
\usepackage{array}
\usepackage{float}
\usepackage{color}
\usepackage{xcolor}
\usepackage{soul}
\usepackage{ulem}
\usepackage{makecell}
\usepackage{hyperref}
\hypersetup{colorlinks=true,linkcolor=blue,anchorcolor=blue,citecolor=red, filecolor=blue,urlcolor=blue,bookmarksnumbered=true,
pdfview=FitB
}

\def\dd{{\mathrm{d}}}

\mathchardef\-="2D

\colorlet{darkgreen}{green!60!black}
\colorlet{brightyellow}{yellow!75!red}
\colorlet{orange}{red!50!yellow}
\colorlet{darkblue}{blue!60!black}
\colorlet{darkred}{red!80!black}
\colorlet{greenblue}{green!50!blue}

\makeatletter

\newcommand{\Rmnum}[1]{\expandafter\@slowromancap\romannumeral #1@}
\makeatother

\begin{document}
\title{Dissecting a strongly coupled scalar nucleon}


\author{Xianghui Cao}
\affiliation{Department of Modern Physics, University of Science \& Technology of China, Hefei 230026, China}

\author{Yang Li}
\thanks{Corresponding author}
\affiliation{Department of Modern Physics, University of Science \& Technology of China, Hefei 230026, China}
\affiliation{Anhui Center for Fundamental Sciences (Theoretical Physics), University of Science and Technology of China, Hefei, 230026, China}

\author{James P. Vary}
\affiliation{Department of Physics and Astronomy, Iowa State University, Ames, IA 50011, USA}

\date{\today}
 \begin{abstract}
We continue our investigation of the stress within a strongly coupled scalar nucleon, and now dissect the gravitational form factors into contributions from its constituents, the (mock) nucleon and the (mock) pion. The computation is based on a non-perturbative solution of the scalar Yukawa model in the light-front Hamiltonian formalism with a Fock sector expansion including up to one nucleon and two pions. By employing the ``good currents"  $T^{++}_i$, $T^{+-}_i$ and $T^{12}_i$, we extract the full set of gravitational form factors $A_i$, $D_i$, $\bar c_i$ without the contamination of the spurious form factors, and free of uncanceled UV divergences. 
With these results, we decompose the mass of the system into its constituents and compute the matter and mechanical radii, gaining insights into the strongly coupled system. 
 \end{abstract}

\maketitle

\section{Introduction}

The hadronic energy-momentum tensor (EMT) provides crucial insight of the system \cite{Polyakov:2018zvc}. On the one hand, it describes the energy and stress distributions within the hadron. On the other hand, it is related to the 3-dimensional partonic structures of the system through the sum rules pioneered by Ji \cite{Ji:1996ek, Ji:1997gm}, 
\begin{equation}
\int \dd x\, xH_i(x, \xi, Q^2) = A_i(Q^2) + \xi^2 D_i(Q^2),
\end{equation}
where, $H_i$ is generalized parton distribution (GPD), $A_i$ and $D_i$ are the gravitational form factors (GFFs), and $i = u, d, g$ refers to each particle species. The latter point is particularly interesting because it provides access to the individual quark and gluon distributions. 
Formally, the individual GFFs can be obtained from the Lorentz decomposition of the hadronic matrix element (HME). For a scalar, such as the pion, the decomposition reads \cite{Polyakov:2018zvc}, 
\begin{multline}\label{eqn:Lorentz_decomposition_HME}
\langle p' | T^{\alpha\beta}_i(0)|p\rangle_\mu 
= \frac{1}{2}(q^\alpha q^\beta - q^2 g^{\alpha\beta})D_i(-q^2, \mu) \\
+ 2 P^\alpha P^\beta A_i(-q^2, \mu) 
+ 2M^2 g^{\alpha\beta} \bar c_i(-q^2, \mu)
\end{multline}
where, $P = \frac{1}{2}(p'+p)$, and $q = p'-p$, $Q^2 = -q^2$. $A_i(-q^2, \mu)$, $D_i(-q^2, \mu)$ and $\bar c_i(-q^2, \mu)$ are the GFFs as mentioned. $i$ refers to specific particle species, e.g. quark. $\mu$ represents the renormalization scale, i.e. the hadronic EMT are renormalization scheme and scale dependent. Unless elsewhere stated, we will suppress the dependence on $\mu$ hereafter for simplicity. 

The GFFs are related to the energy and stress densities of the system; however their exact physical interpretation is under debate \cite{Lorce:2017xzd, Lorce:2018egm, Ji:2021qgo, Ji:2021mtz, Lorce:2021xku, Ji:2021znw, Freese:2021czn, Freese:2021qtb,  Freese:2021mzg, Ji:2022exr, Freese:2022fat, Freese:2023abr, Lorce:2024ipy, Li:2024vgv}. Lorcé proposed that relativistic fluidity can be used to define the physical densities of a hadron \cite{Lorce:2017xzd, Teryaev:2022pke}. Following this lead,
we have established that the hadronic EMT $\mathcal T^{\alpha\beta}$ can be written in a form closely resembling the EMT of a relativistic continuum \cite{Li:2024vgv}, 
\begin{equation}\label{eqn:continuum_EMT}
\mathcal T^{\alpha\beta} = e u^\alpha u^\beta - p \Delta^{\alpha\beta} + \pi^{\alpha\beta}
\end{equation}
where, $u^\alpha$ is the comoving velocity, $\Delta^{\alpha\beta} = g^{\alpha\beta} - u^\alpha u^\beta$ is the spatial metric tensor.  $e$, $p$ and $\pi^{\alpha\beta}$ are the proper energy density, pressure and shear stress, respectively. 
These physical densities are related to the linear combination of GFFs by Fourier transform \cite{Li:2024vgv}. In particular, $\pi^{\alpha\beta}$, the non-dissipative shear tensor, distinguishes hadrons from the ideal fluid \cite{Rezzolla:2013dea}. 
The hadronic EMT $\mathcal T^{\alpha\beta}$ is obtained by deconvoluting the wavepacket from the hadronic quantum expectation value \cite{Li:2022hyf, Li:2024vgv}. 
The above formalism can be generalized to the individual EMT, 
\begin{equation}\label{eqn:continuum_EMT_decomposition}
\mathcal T^{\alpha\beta}_i = e_i u^\alpha u^\beta - p_i \Delta^{\alpha\beta} + \pi^{\alpha\beta}_i + g^{\alpha\beta} \Lambda_i,
\end{equation}
where, $e_i$, $p_i$ and $\pi^{\alpha\beta}_i$ are the physical densities associated with an individual constituent. 
A new density $\Lambda_i$, in analogy to cosmological constant \cite{Teryaev:2013qba, Teryaev:2016edw, Liu:2023cse}, emerges due to the non-conserving nature of the individual EMT $\mathcal T^{\alpha\beta}_i$, i.e. EMT of each particle species. It represents the external pressure acting on the $i$-th constituent, and is related to $\bar c_i$ by Fourier transform. 
The physical densities defined in (\ref{eqn:continuum_EMT}) are the sum of the individual densities, e.g., the energy density,
$e = \sum_i e_i$.
In hydrodynamics, such a decomposition is known as the \textit{coupled multi-fluid picture} \cite{Rezzolla:2013dea}. In this picture, all fluid species have the same four-velocity, which is in contrast to the interacting multi-fluid picture. The latter is more suitable for describing loosely bound hadrons, e.g.  hadron molecules and nuclei \cite{He:2024jgc}. 

Since the energy density $e(x)$ is normalized to the hadron mass $M$, dissection of this quantity into constituents provides a quantitative picture of the origin of the hadron mass. It is customary to include the external pressure $\Lambda_i$ as part of the individual contribution to the mass, viz.~$U_i \equiv  u_\alpha u_\beta \mathcal T^{\alpha\beta}_i$. Upon integrating over the entire space, the fractional contribution of the $i$-th constituent to the total hadron mass is $f_i = A_i(0) + \bar c_i(0)$ \cite{Lorce:2017xzd, Lorce:2021xku}. For the proton, lattice simulations estimate that 43\% of its mass originates from the quarks and 57\% from the gluons at 2 GeV within the D2 scheme up to 3-loops \cite{Metz:2020vxd}. 
The decomposition of the spin and the pressure are conceptually similar and are also investigated in the literature \cite{Burkert:2023wzr}.  

In this work, we investigate the decomposition of the hadronic EMT within a (3+1)D strongly coupled scalar theory. This is a continuation of our previous work on the evaluation of the GFFs from the same theory \cite{Cao:2023ohj}. The simplicity of the scalar theory allows us to apply the light-front Hamiltonian approach with a systematic Fock space expansion even in the strong-coupling regime \cite{Li:2014kfa, Li:2015iaw, Karmanov:2016yzu}. The method provides access to the microscopic description of the GFFs as well as the related physical densities with a wave function representation, in analogy to the famous Drell-Yan-West formula for the electromagnetic form factors \cite{Drell:1969km, West:1970av, Brodsky:1980zm, Brodsky:1998hn, Brodsky:2000ii}. 
One of the main challenges to compute the GFFs within the light-front Hamiltonian formalism is the non-perturbative renormalization of the energy-momentum tensor operator. Since the light-front formulation is not manifestly covariant, the divergences of different components of the operator could be dramatically different.  
To extract GFFs $A(Q^2)$ and $D(Q^2)$, we employed operators $T^{++}$ and $T^{+-}$ \cite{Cao:2023ohj}. Here, we adopt light-front coordinates, $x^\pm = x^0 \pm x^3$ and $\vec x_\perp = (x^1, x^2)$. In particular, we showed that $T^{+-}$ is properly renormalized using the counterterms for the light-front Hamiltonian $P^-$. Indeed, $T^{+-}$ is the density of the light-front Hamiltonian \cite{Brodsky:1997de}. 
To access the individual GFFs $A_i(Q^2)$, $D_i(Q^2)$ and $\bar c_i(Q^2)$, we need one more current. One of the natural choices is $T^{11}_i$ (or $T^{22}_i$ or their sum); however one can show that the HME is divergent, even for the total current $T^{11}$. Note that using the divergence of the $T^{\mu 1}_i$ is equivalent to adopting $T^{11}_i$.  
In Ref.~\cite{Cao:2024rul}, we performed a systematic covariant analysis of the HME of the light-front EMT based on a comparison of the covariant and light-front perturbation theory. We found that $T^{11}_i$ and $T^{22}_i$ are contaminated by the so-called spurious form factors, which contain uncanceled UV divergences. We also concluded that $T_i^{++}$, $T^{+-}_i$ and $T_i^{12}$ are ``good currents" that are free of the spurious form factors and can be used to extract the full set of GFFs. In this work, we apply this formulation to the scalar Yukawa theory in the non-perturbative regime. 

The remainder of the work is organized as follows. Sect.~\ref{sect:HME_CLFD} introduces the covariant analysis of the HME on the light front, and identifies the good currents $T^{++}_i$, $T^{+-}_i$ and $T_i^{12}$ that can be used to extract the GFFs. We evaluate the HME in Sect.~\ref{sect:wave_function_representation} and present the wave function representation of the HME of the good currents. 
The numerical results are shown in Sect.~\ref{sect:numerics}. We conclude in Sect.~\ref{sect:summary}. 

\section{Covariant analysis of hadronic matrix element on the light front}\label{sect:HME_CLFD}

The Lagrangian of the scalar Yukawa model reads \cite{Li:2014kfa, Li:2015iaw, Karmanov:2016yzu}, 
 \begin{multline}
 \mathscr L = \partial_\mu N^\dagger \partial^\mu N - m^2 N^\dagger N 
 + \frac{1}{2}\partial_\mu \pi \partial^\mu \pi - \frac{1}{2}\mu^2 \pi^2 \\
 + g_0 N^\dagger N \pi + \delta m^2 N^\dagger N
 + \frac{1}{2} \delta \mu^2 \pi^2 \,,
 \end{multline}
 where, $\delta m^2$ and $\delta \mu^2$ are mass counterterms and $g_0$ is the bare coupling.  These parameters can be obtained from the mass and coupling constant renormalization. This Lagrangian describes the interaction between a complex scalar field $N$ and a real scalar field $\pi$. We assign the nucleon mass $m = 0.94\,\mathrm{GeV}$ and the pion mass $\mu = 0.14\,\mathrm{GeV}$ to the physical mass of these two fields, respectively.  Note that the physical coupling $g$ is dimensional. Hence, the theory is super-renormalizable. Nevertheless, ultraviolet divergences still exist. To regularize these divergences, we introduce the Pauli-Villars scheme, which works well for non-perturbative quantum field theories on the light cone \cite{Chabysheva:2010vk, Hiller:2016itl}. Unless otherwise stated, we will suppress the notations associated with the Pauli-Villars particles and refer to Refs.~\cite{Li:2015iaw, Karmanov:2016yzu} for the details. 

The EMT of the scalar Yukawa theory is, 
 \begin{multline}
 T^{\mu\nu} = \partial^{\{\mu}N^\dagger \partial^{\nu\}}N - g^{\mu\nu} (\partial_\rho N^\dagger \partial^\rho N - m^2 N^\dagger N) \\
 - g^{\mu\nu} g_0 N^\dagger N \pi - g^{\mu\nu} \delta m^2 N^\dagger N 
 + \partial^\mu \pi \partial^\nu \pi  \\- \frac{1}{2} g^{\mu\nu} (\partial_\sigma \pi \partial^\sigma \pi - \mu^2 \pi^2) 
 - \frac{1}{2}g^{\mu\nu} \delta \mu^2 \pi^2\,.
 \end{multline}
 Here, $a^{\{\mu}b^{\nu\}} \equiv a^\mu b^\nu + a^\nu b^\mu$. 
Following the mass decomposition of hadrons \cite{Ji:1994av, Ji:1995sv}, we split the EMT of the scalar Yukawa theory into a ``material part" and a ``radiation part", viz. $T^{\mu\nu} = T_N^{\mu\nu} + T_\pi^{\mu\nu}$. The former includes the free nucleon EMT as well as the interaction, and the latter is simply the free pion EMT,  
\begin{align}
    T_N^{\mu\nu} =\,&  \partial^{\{\mu}N^\dagger \partial^{\nu\}}N  
     - g^{\mu\nu}\big(\partial_\sigma N^\dagger\partial^\sigma N - m^2 N^\dagger N \big) \nonumber \\
    & - g^{\mu\nu} g N^\dagger N \pi,  \label{eqn:EMT_N} \\
    T_\pi^{\mu\nu} =\,& \partial^\mu\pi \partial^\nu\pi - \frac{1}{2}g^{\mu\nu} \big(\partial^\rho\pi\partial_\rho\pi - \mu^2\pi^2\big).   \label{eqn:EMT_pi} 
\end{align}
The Lorentz decomposition of the corresponding HME is given by Eq.~(\ref{eqn:Lorentz_decomposition_HME}). 
Summing over all constituents gives the total GFFs, 
\begin{align}
 \sum_i A_i(-q^2, \mu) =\,& A(-q^2), \\ 
\sum_i D_i(-q^2, \mu) =\,& D(-q^2), \\  
 \sum_i \bar c_i(-q^2, \mu) = \,& \bar c(-q^2).
\end{align}
GFF $A_i$ can be interpreted as the matter (number) density and $D_i$ is related to the internal stress. 
GFF $\bar c_i$ represents the external stress, i.e. the stress between the $i$-th component and the rest of the system. 
Current conservation requires that the total force vanishes for a self-bound system, viz. $\bar c(Q^2) = 0$.

The Lorentz decomposition of the HME (\ref{eqn:Lorentz_decomposition_HME}) is based on the Poincaré symmetry. In practical  calculations in the non-perturbative regime, the Poincaré symmetry may not be fully preserved. One of the advantages of light-front dynamics is that it retains the maximal number of kinematical symmetries \cite{Dirac:1949cp}. 
Karmanov et. al. proposed covariant light-front dynamics (CLFD) to reparametrize the HMEs in terms of the kinematical subgroup of the Poincaré group on the light front \cite{Carbonell:1998rj}. For the EMT, one of the possible parametrization of the HME within the Drell-Yan frame $q^+=0$ is \cite{Cao:2024rul}, 
\begin{widetext}
\begin{multline}\label{eqn:CLFD_decomposition}
    \langle p' | T_{i}^{\alpha\beta}(0) | p\rangle = 2P^\alpha P^\beta A_i(-q^2) + \frac{1}{2}(q^\alpha q^\beta - q^2 g^{\alpha\beta}) D_i(-q^2) 
    +  2M^2 g^{\alpha\beta}\bar{c}_i(-q^2) \\
    + \frac{M^4 \omega^\alpha \omega^\beta}{(\omega\cdot P)^2}S_{1i}(-q^2) + (V^\alpha V^\beta + q^\alpha q^\beta) S_{2i}(-q^2),
\end{multline}
\end{widetext}
where, $P = (p'+p)/2$, $q = p'-p$, and $M^2 = p^2 = p'^2$. $\omega^\mu = (\omega^+, \omega^-, \vec\omega_\perp) = (0, 2, 0)$ is a null vector indicating the orientation of the quantization surface, the light front $\Sigma = \{x\in \mathbb R^{3,1}|\omega\cdot x = 0\}$, and $V^\alpha = \varepsilon^{\alpha \beta \rho \sigma}P_\beta q_\rho \omega_\sigma/(\omega\cdot P)$. Vector $V^\alpha$ is perpendicular to $q^\alpha$, $P^\alpha$ as well as $\omega^\alpha$. 
Note that two more covariant structures emerge reflecting the reduction of the symmetry. The corresponding form factors $S_{1i}$ and $S_{2i}$ are known as spurious form factors. Namely, they are artifacts of the formulation and should vanish in the physical limit when the full Poincaré symmetry is restored. 
Hermiticity excludes terms like $P^{\{\alpha} q^{\beta\}}$, $\omega^{\{\alpha} q^{\beta\}}$ or $\omega^{\{\alpha} V^{\beta\}}$. 
The parametrization (\ref{eqn:CLFD_decomposition}) is not unique in CLFD \cite{Cao:2023ohj}. In Ref.~\cite{Cao:2024rul}, we have shown, by comparing results from covariant 
perturbation theory and light-front perturbation theory,  that Eq.~(\ref{eqn:CLFD_decomposition}) removes the UV divergences in the physical form factors $A$, $D$, $\bar c$. Furthermore, it is also consistent with calculation from light-cone perturbation theory (LCPT). The underlying reason is that it endows a rotation symmetry within the transverse plane within the spurious terms. In this work, we therefore adopt this parametrization (\ref{eqn:CLFD_decomposition}) as our starting point for non-perturbative calculation. 

From Eq.~(\ref{eqn:CLFD_decomposition}), the HMEs of the good current $T_i^{++}$, $T^{+-}_i$ and $T_i^{12}$ are, 
\begin{align}
 t^{++}_i=\,& 2(P^{+})^2 A_i(q^2_\perp), \label{eqn:t++_CLFD_breit}\\
t^{12}_i =\,& \frac{1}{2} q^1_\perp q^2_\perp  D_i(q^2_\perp) \label{eqn:t12_CLFD_breit} \\
 t^{+-}_i =\,& 2(M^2+\frac{1}{4}q_\perp^2)A_i(q^2_\perp) + q^2_\perp D_i(q^2_\perp) \nonumber \\
& + 4M^2\bar c_i(q^2_\perp). \label{eqn:t+-_CLFD_breit}
\end{align}
Here, we have adopted the Drell-Yan-Breit frame ($q^+ = 0, \vec P_\perp = 0$) to simplify the algebra. 
Note that these HMEs are not  contaminated by the spurious GFFs as desired. 
By contrast, one can show from (\ref{eqn:CLFD_decomposition}) that HME of $T^{11}+T^{22}$ contains spurious form factor $S_{2i}$, 
\begin{multline}
t^{11}_i + t^{22}_i =  - \frac{1}{2} q^2_\perp D_i(q^2_\perp) 
- 4M^2 \bar c_i(q^2_\perp) \\
+ 2q_\perp^2S_{2i}(q_\perp^2).
\end{multline}
The total spurious form factor $S_{2} = \sum_i S_{2i}$ is non-vanishing and UV divergent.

\section{Wave function representation} \label{sect:wave_function_representation}

\begin{figure*}
    \centering
    \includegraphics[width=0.23\textwidth]{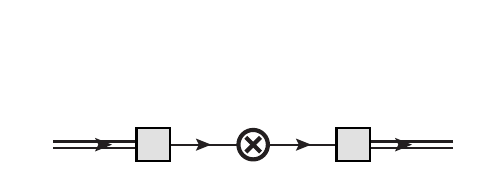}
\includegraphics[width=0.23\textwidth]{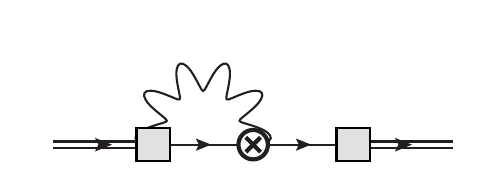}
\includegraphics[width=0.23\textwidth]{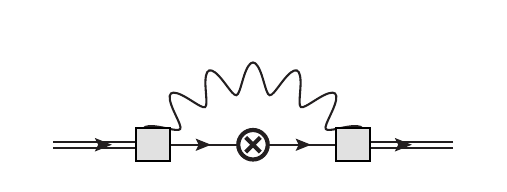}
\includegraphics[width=0.23\textwidth]{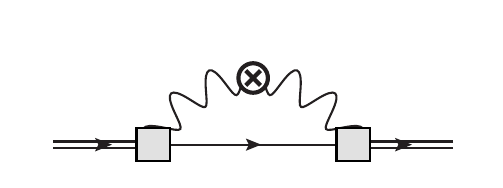}
\\
\includegraphics[width=0.23\textwidth]{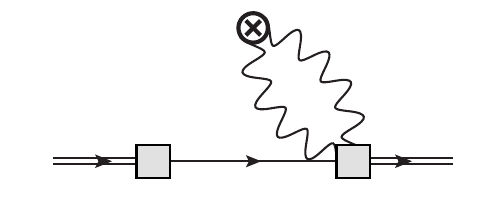}
\includegraphics[width=0.23\textwidth]{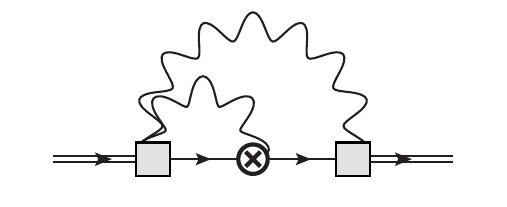}
\includegraphics[width=0.23\textwidth]{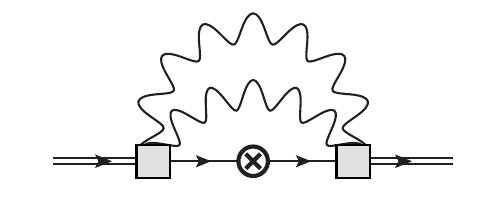}
\includegraphics[width=0.23\textwidth]{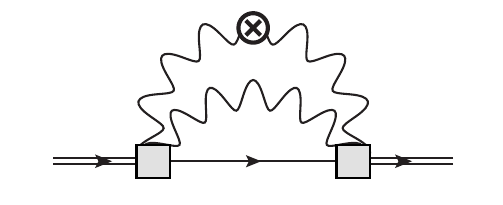}
    \caption{Diagrammatic representations of the matrix elements of the EMT. See Ref.~{\cite{Cao:2023ohj}} for more details on the evaluation of these diagrams. }
    \label{fig:emt_ff}
\end{figure*}

The HME $t^{\alpha\beta}_i$ were previously evaluated up to three-body truncation (one scalar nucleon plus two scalar pions) in the non-perturbative regime in Ref.~\cite{Cao:2023ohj}. 
The relevant diagrammatic representation is shown in Fig.~\ref{fig:emt_ff}.  The non-perturbative wave functions (shaded boxes) were obtained in our previous work with a Fock space expansion up to three-body $|N\rangle+|N\pi\rangle+|N\pi\pi\rangle$. A further solution up to four-body $|N\rangle+|N\pi\rangle+|N\pi\pi\rangle+|N\pi\pi\pi\rangle$  showed that the three-body truncation achieved numerical convergence \cite{Li:2014kfa, Li:2015iaw, Karmanov:2016yzu, Duan:2024dhy}.  Summing up contributions from these diagrams, we obtain the wave function representation of the HMEs, which can be straightforwardly generalized to the $n$-body case. The wave function representation for $t_i^{++}$, $t^{+-}_i$ and $t^{12}_i$ are, 
    \begin{align}
        t^{++}_i =\,& \sum_n \int \big[ \mathrm{d}x_k \mathrm{d}^2 r_{k\perp}  \big]_n  \widetilde{\psi}_n^*(\{ x_k, \vec{r}_{k\perp} \})  \label{eqn:t++_WFR} \\
       & \times \sum_j x_j  e^{i\vec{r}_{j\perp} \cdot \vec{q}_\perp} \widetilde{\psi}_n(\{x_k, \vec{r}_{k\perp}\}),  \nonumber \\
        t^{+-}_i =\,& 2 \sum_n \int \big[ \mathrm{d}x_k \mathrm{d}^2 r_{k\perp}  \big]_n \widetilde{\psi}_n^*(\{ x_k, \vec{r}_{k\perp} \}) \label{eqn:t+-_WFR} \\
        & \times \sum_j e^{i\vec{r}_{j\perp} \cdot \vec{q}_\perp}  \bigg( \frac{- \frac{1}{4} \tensor{\nabla}^2_{j\perp} + m_j^2 - \frac{1}{4}q_\perp^2}{x_j}\bigg) \nonumber \\
       &  \times \widetilde{\psi}_n(\{x_k, \vec{r}_{k\perp}\}) \nonumber \\
        -\,&  2 \sum_n \int\big[ \mathrm{d}x_k \mathrm{d}^2 r_{k\perp}  \big]_n \widetilde{\psi}_n^*(\{ x_k, \vec{r}_{k\perp} \}) \nonumber \\
        & \times e^{i \vec{r}_{n\perp} \cdot \vec{q}_\perp} 
        \bigg( \sum_j \frac{- \vec{\nabla}_{j\perp}^2 + m_j^2}{x_j} - M^2 \bigg) \nonumber \\
        &\times \widetilde{\psi}_n(\{ x_k, \vec{r}_{k\perp} \}), \nonumber \\
t^{12}_i =\,& \frac{1}{2} \sum_n \int \big[\dd x_k \dd^2 r_{k\perp}\big]_n \widetilde\psi_n^*\big(\{x_k, \vec r_{k\perp}\}\big) \label{eqn:t12_WFR} \\
& \times \sum_j e^{-i\vec q_\perp \cdot \vec r_{j\perp}} 
 \frac{i\tensor\nabla_{j1}i\tensor\nabla_{j2} - q_1 q_2}{x_j} \nonumber \\
 & \times \widetilde\psi_n\big(\{x_k, \vec r_{k\perp}\}\big),    \nonumber    
    \end{align}
where, $f \tensor \nabla g = f (\vec\nabla g) - (\vec \nabla f) g$, and $j$ sums over all partons belongs to the constituent of the $i$-th type. $\big[ \mathrm{d}x_k \mathrm{d}^2 r_{k\perp}  \big]_n$ is the $n$-body Fock space measure. We denote the $n$-body LFWFs as $\widetilde{\psi}_n(\{ x_k, \vec{r}_{k\perp} \})$ with the argument indicating the required set of constituent coordinates. See Ref.~{\cite{Cao:2023ohj}} for more details.  Note that non-perturbative renormalization has been taken into account in these remarkably simple expressions, similar to the wave function representation for the electroweak current \cite{Brodsky:1998hn, Brodsky:2000ii}. The expressions for $t^{++}_i$ and $t^{+-}_i$ are previously obtained in Ref.~\cite{Cao:2023ohj}. Here we correct a typographical error for $t^{+-}_i$. 
It is tempting to extend the expression (\ref{eqn:t12_WFR}) to the full transverse stress tensor $t^{ab}_i$ where $a=1,2; b=1,2$. However, the $T^{12}_i$ operator is kinematical whereas the $T^{11}_i$ and $T^{22}_i$ operators are dynamical, and may contain spurious contributions for realistic wave functions. 
From Eq.~(\ref{eqn:t12_WFR}), we obtain a general wave function representation for the Polyakov-Weiss $D$-term $D_i \equiv D_i(0)$, 
\begin{multline}\label{eqn:t12}
D_i = \sum_n \int \big[\dd x_k \dd^2 r_{k\perp}\big]_n   \widetilde\psi_n^*\big(\{x_k, \vec r_{k\perp}\}\big) \\
\times \sum_j 
 \frac{ r_{j1} \tensor\nabla_{j1} r_{j2} \tensor\nabla_{j2} - 1}{x_j}\widetilde\psi_n\big(\{x_k, \vec r_{k\perp}\}\big).
\end{multline}

\section{Numerical results}\label{sect:numerics}

From the HME of the good currents $T^{++}_i$, $T^{+-}_i$ and $T^{12}_i$, we are able to extract the GFFs $A_i$, $D_i$ and $\bar c_i$, where $i = N, \pi$. Note that these quantities are renormalization scheme and scale dependent. In this work, we adopt the Pauli-Villars regularization with a Pauli-Villars mass $\mu_\textsc{pv} = 15\,\mathrm{GeV}$. In LCPT, the Pauli-Villars regularization can be converted to the $\overline{\mathrm{MS}}$ scheme \cite{Cao:2024rul}. However, it is not clear how such a conversion can be done for strongly coupled scalar theory, since the $\overline{\mathrm{MS}}$ scheme is only defined in dimensional regularization in perturbation theory. Nevertheless, the GFFs $A_{i}$ and $D_{i}$  of the scalar Yukawa theory are renormalization scheme independent in the continuum limit ($\mu_\textsc{pv} \to \infty$). 

\begin{figure}
\subfigure[\ \label{fig:GFFs_ADC_LCPT}]{\quad\quad\includegraphics[width=0.4\textwidth]{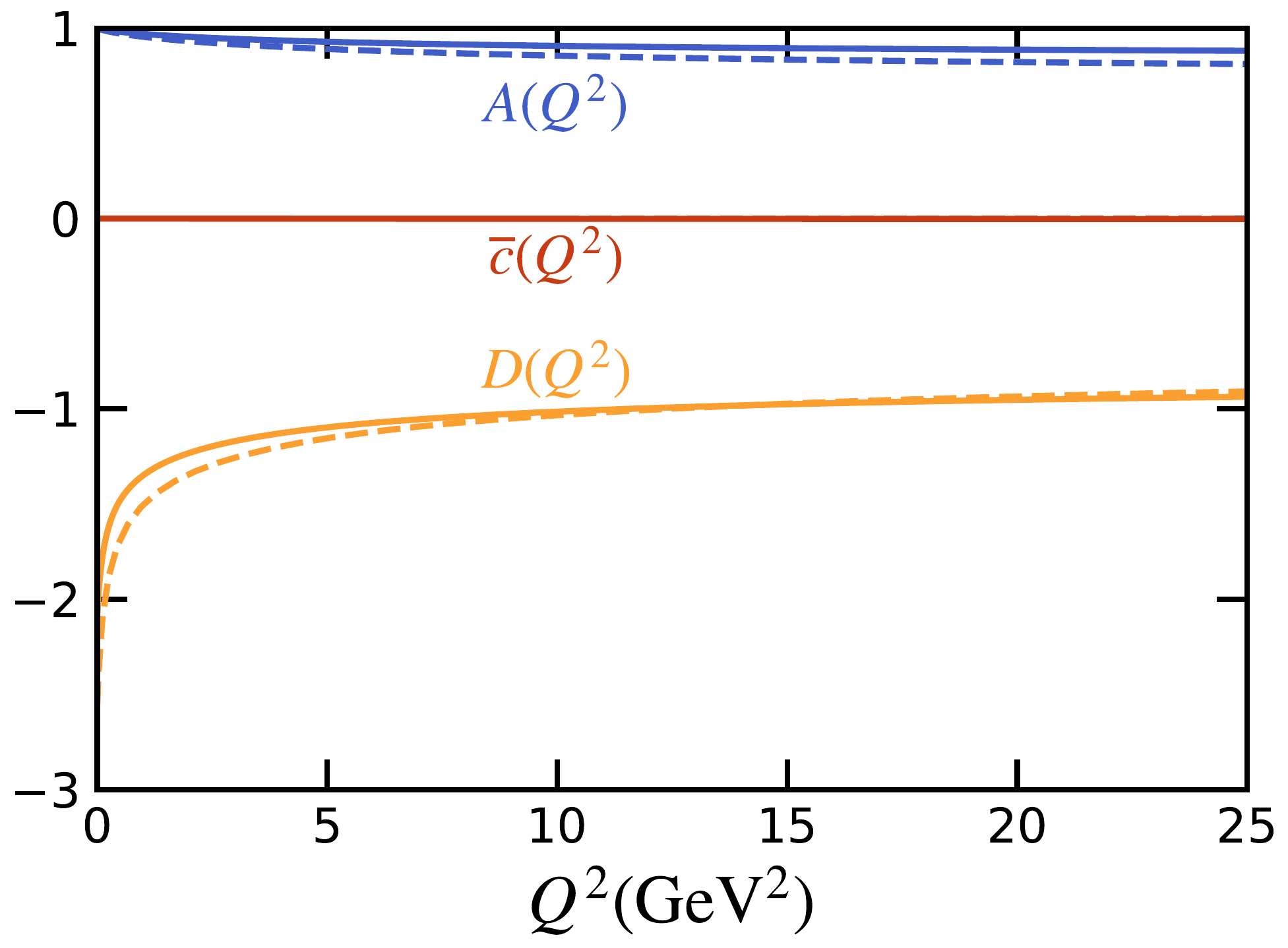}} 
\subfigure[\ \label{fig:GFF_Di_LCPT}]{\includegraphics[width=0.44\textwidth]{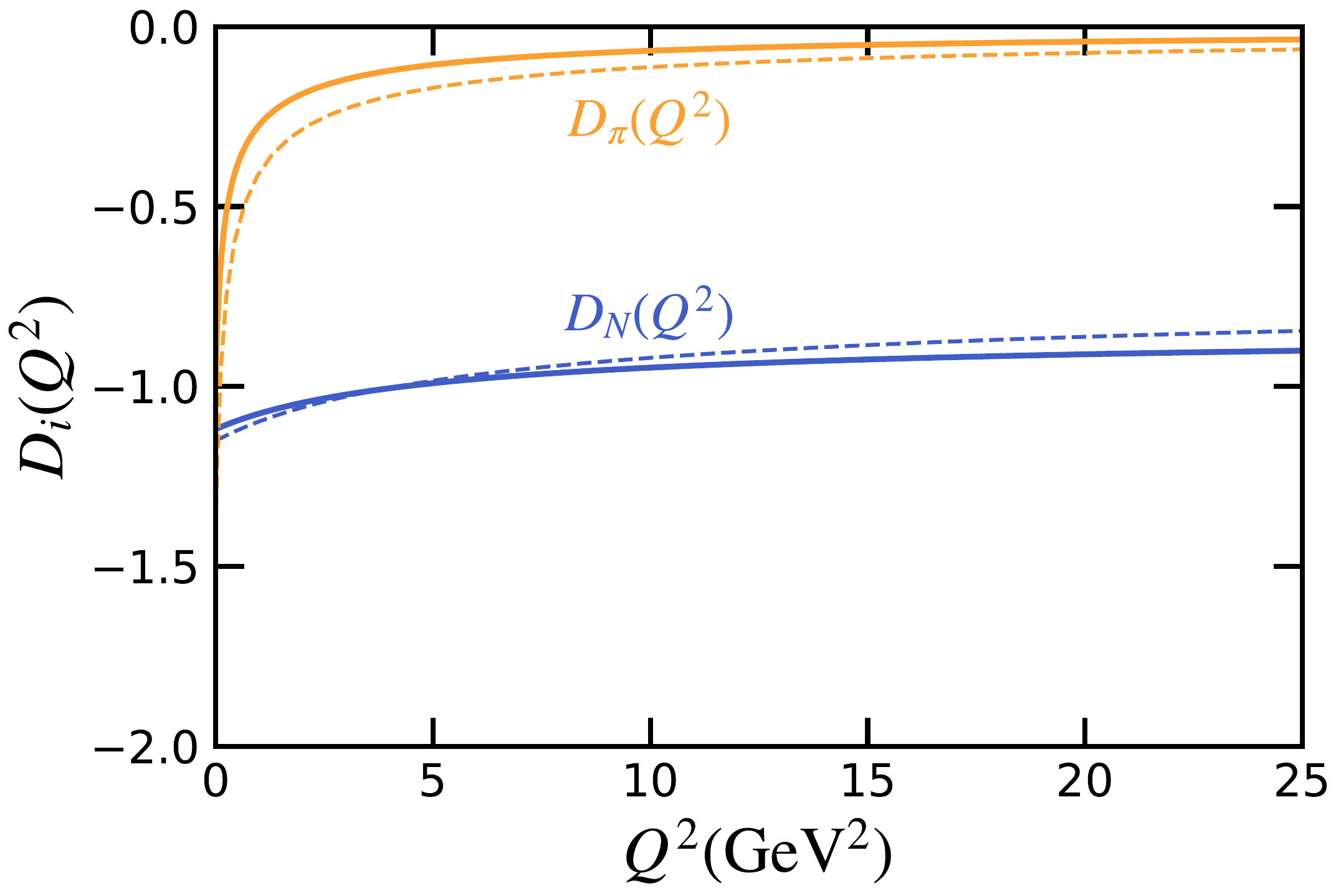}} 
\subfigure[\ \label{fig:GFF_ci_LCPT}]{\includegraphics[width=0.44\textwidth]{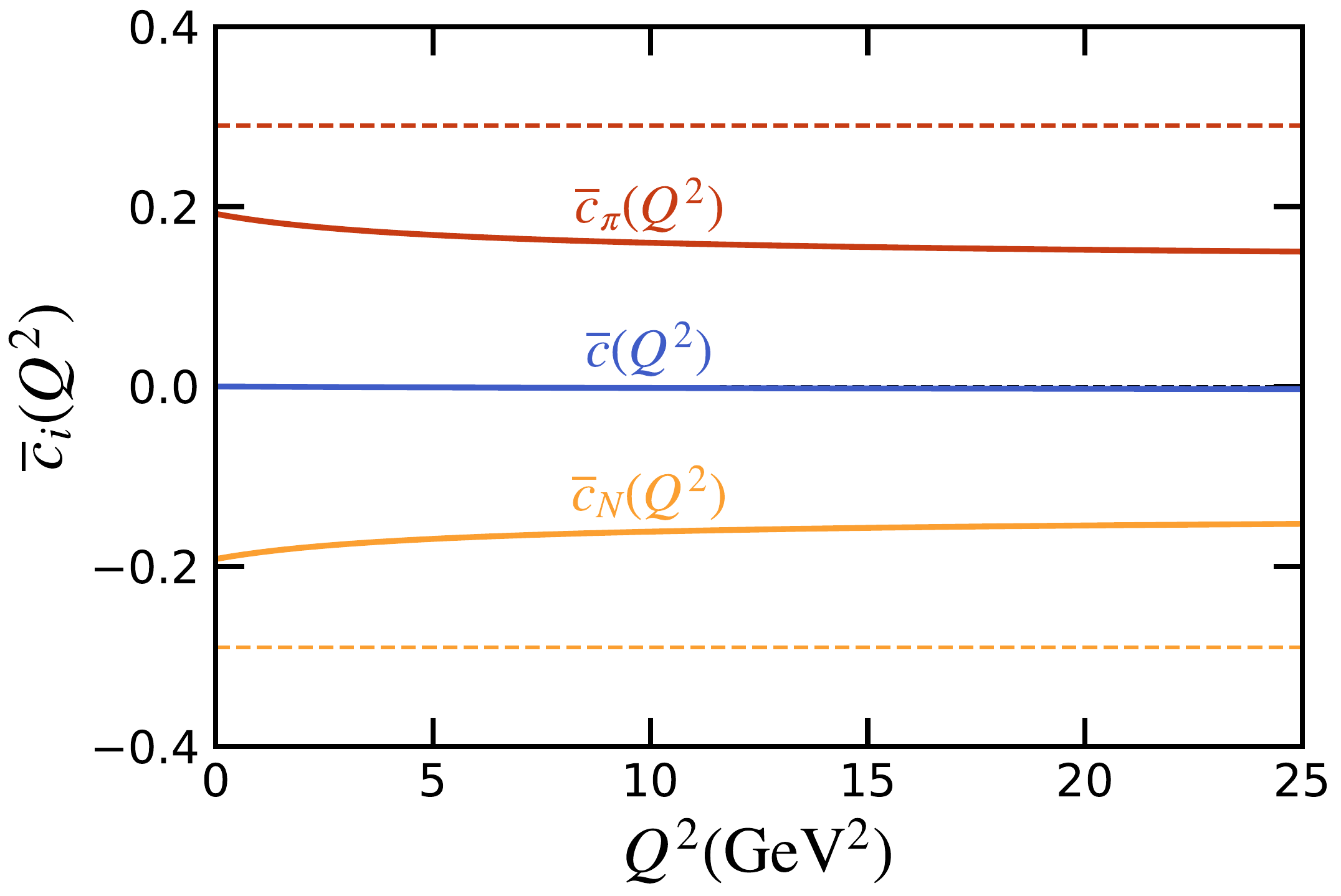}}
\caption{Comparison of the gravitational form factors obtained from light-cone perturbation theory (dashed lines) and from our non-perturbative solution (solid lines) with a three-body truncation (up to one scalar nucleon and two scalar pions). The coupling is $\alpha = 0.5$ and the Pauli-Villars mass $\mu_\textsc{pv} = 15\,\mathrm{GeV}$. }
\label{fig:GFFs_LCPT}
\end{figure}

\begin{figure}
\subfigure[\ \label{fig:GFF_Ai}]{\includegraphics[width=0.42\textwidth]{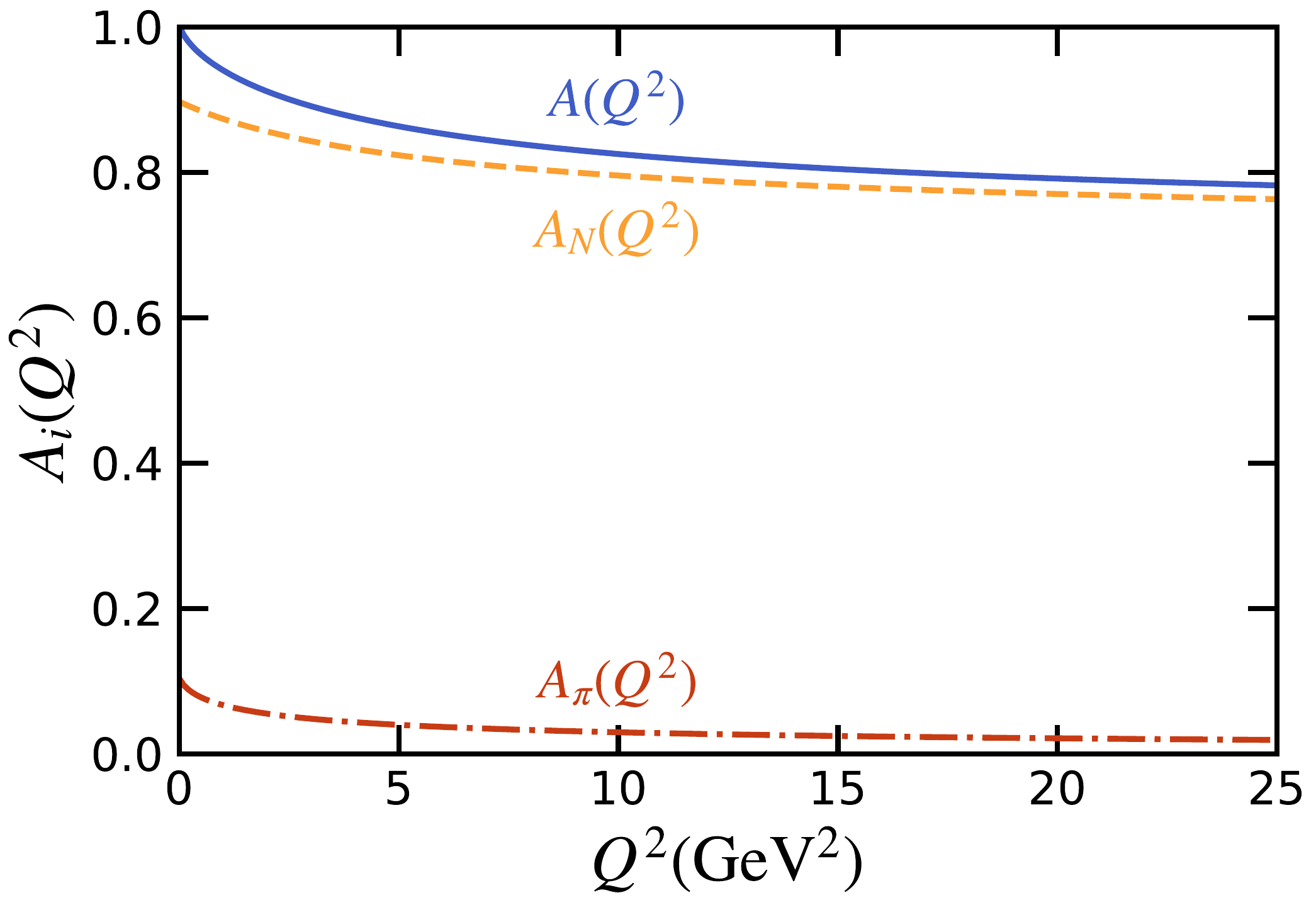}}
\subfigure[\ \label{fig:GFF_Di}]{\includegraphics[width=0.42\textwidth]{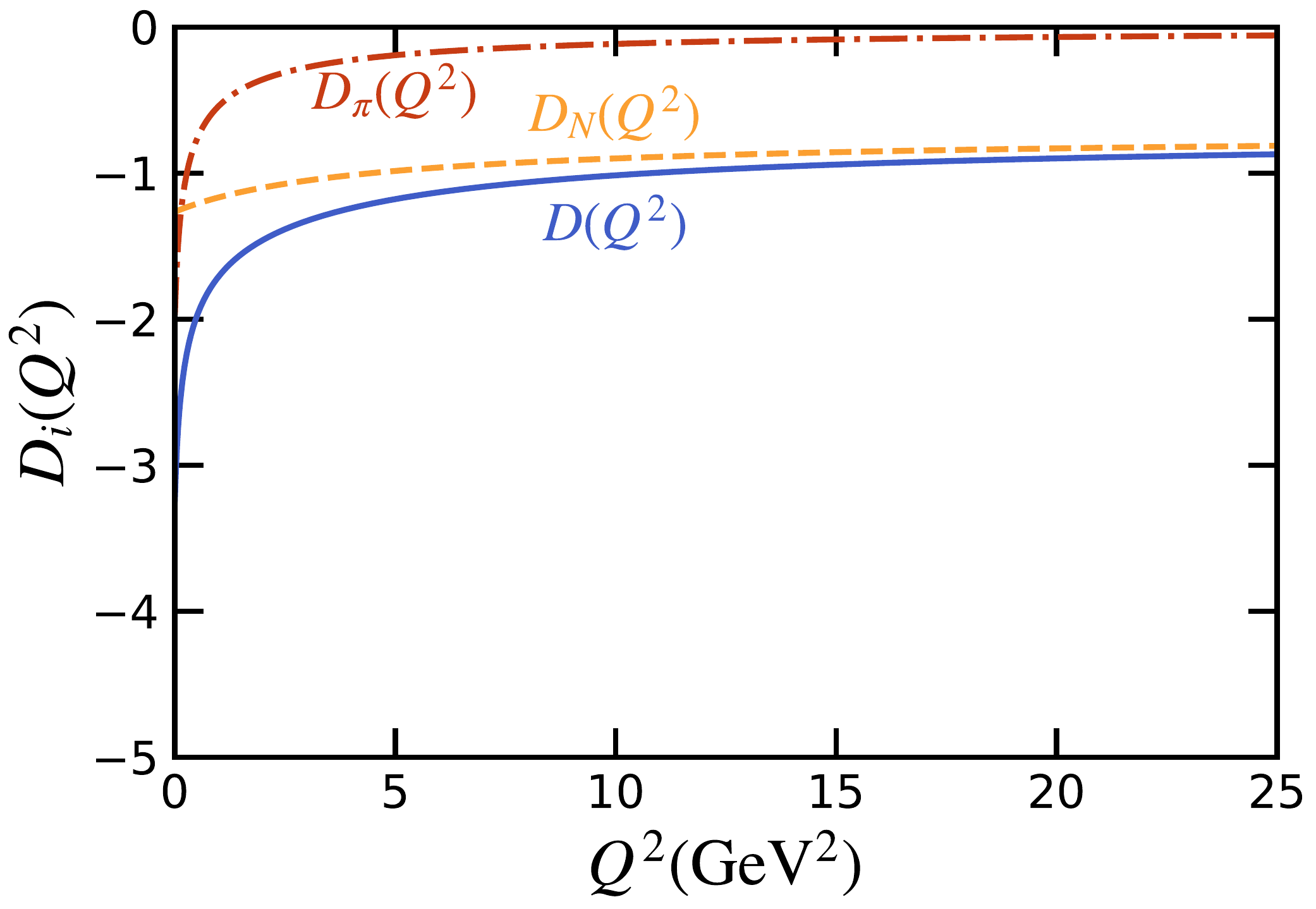}}
\subfigure[\ \label{fig:GFF_ci}]{\includegraphics[width=0.43\textwidth]{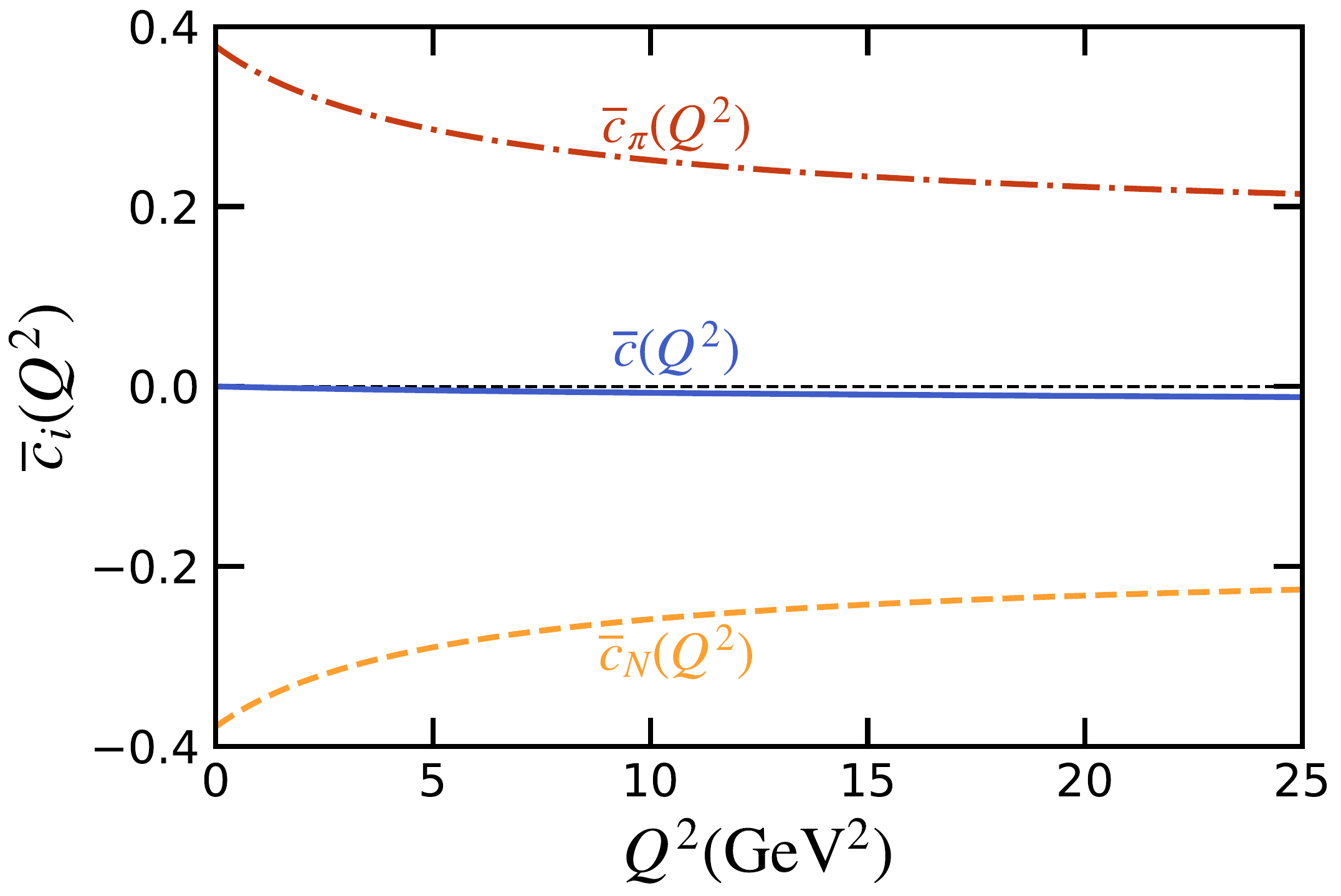}}
\caption{Gravitational form factors $A_i(Q^2)$, $D_i(Q^2)$ and $\bar c_i(Q^2)$ of the physical scalar nucleon, where $i=N, \pi$ for the nucleon (dashed lines) and the pion (dot-dashed lines), respectively. $A(Q^2) = \sum_i A_i(Q^2)$, $D = \sum_i D_i(Q^2)$ and $\bar c(Q^2) = \sum_i \bar c_i (Q^2)$ are the total GFFs summing over all constituents (solid lines). 
The scalar Yukawa theory is solved in a light-front Hamiltonian approach with a non-perturbative coupling $\alpha = 1.0$ and Pauli-Villars mass $\mu_\textsc{pv} = 15\,\mathrm{GeV}$ in a three-body truncation (up to one scalar nucleon and two scalar pions). }
\label{fig:GFFs}
\end{figure}

\begin{figure}
\subfigure[\ \label{fig:GFF_Ai_alpha2p0}]{\includegraphics[width=0.42\textwidth]{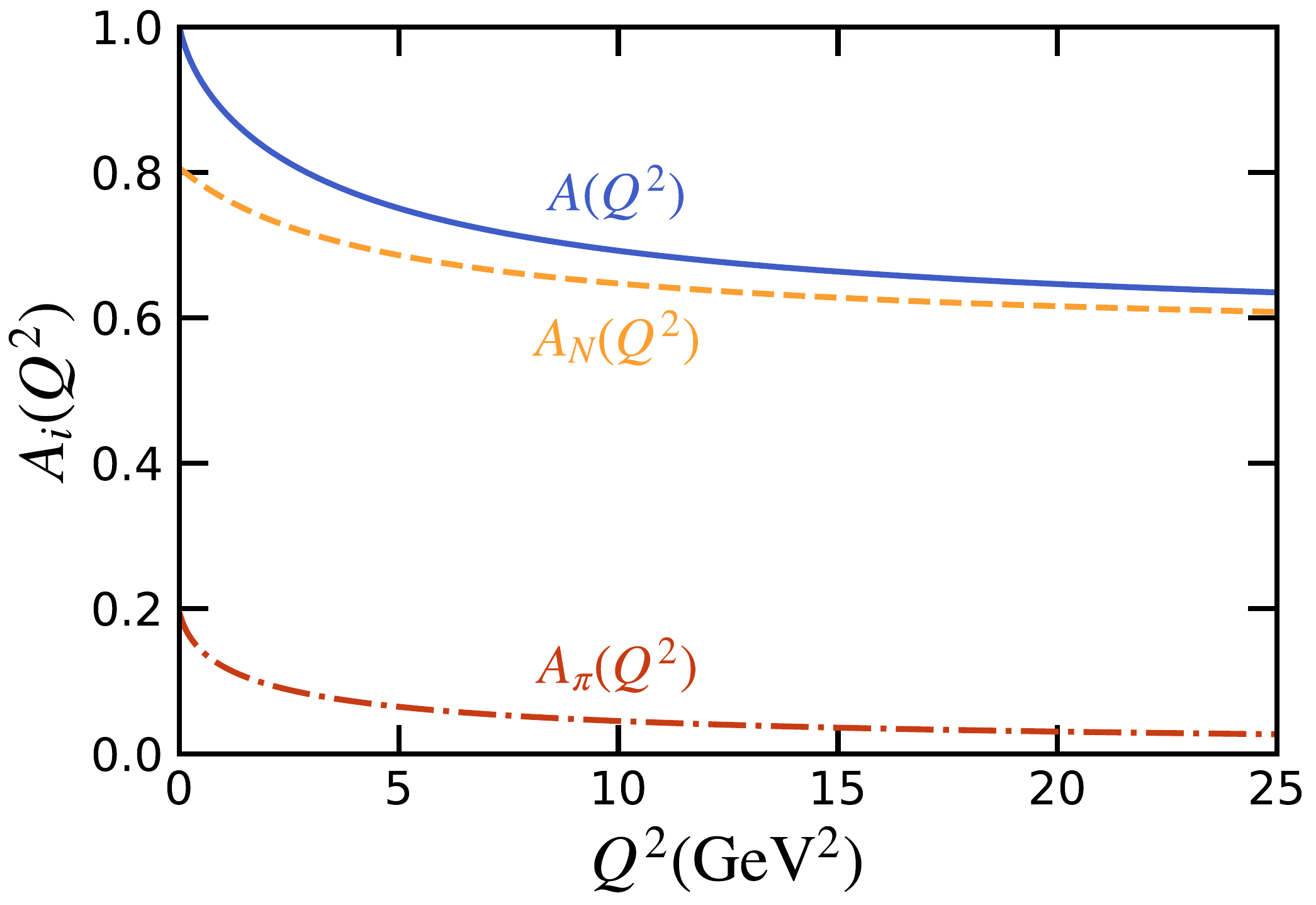}}
\subfigure[\ \label{fig:GFF_Di_alpha2p0}]{\includegraphics[width=0.42\textwidth]{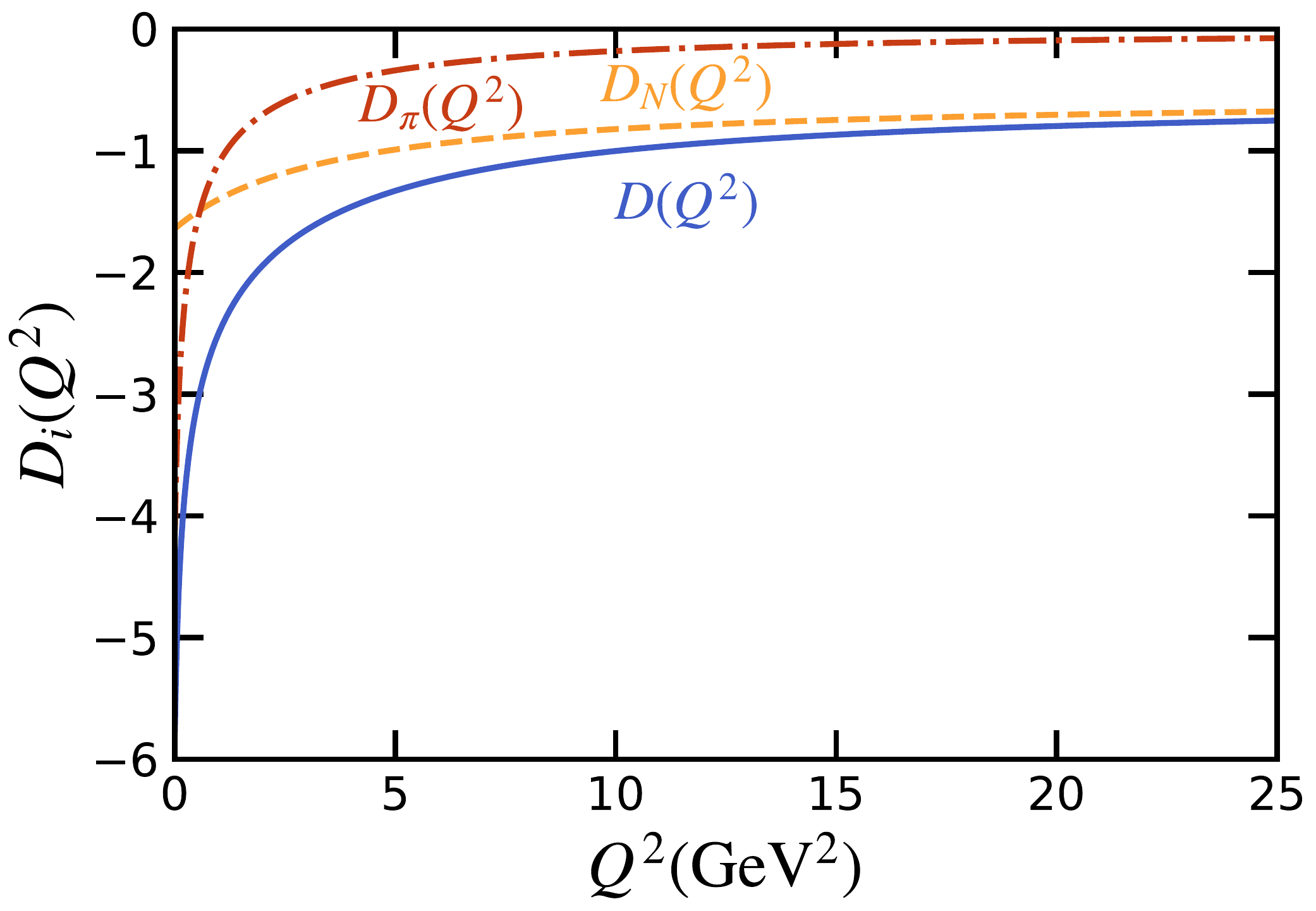}}
\subfigure[\ \label{fig:GFF_ci_alpha2p0}]{\includegraphics[width=0.43\textwidth]{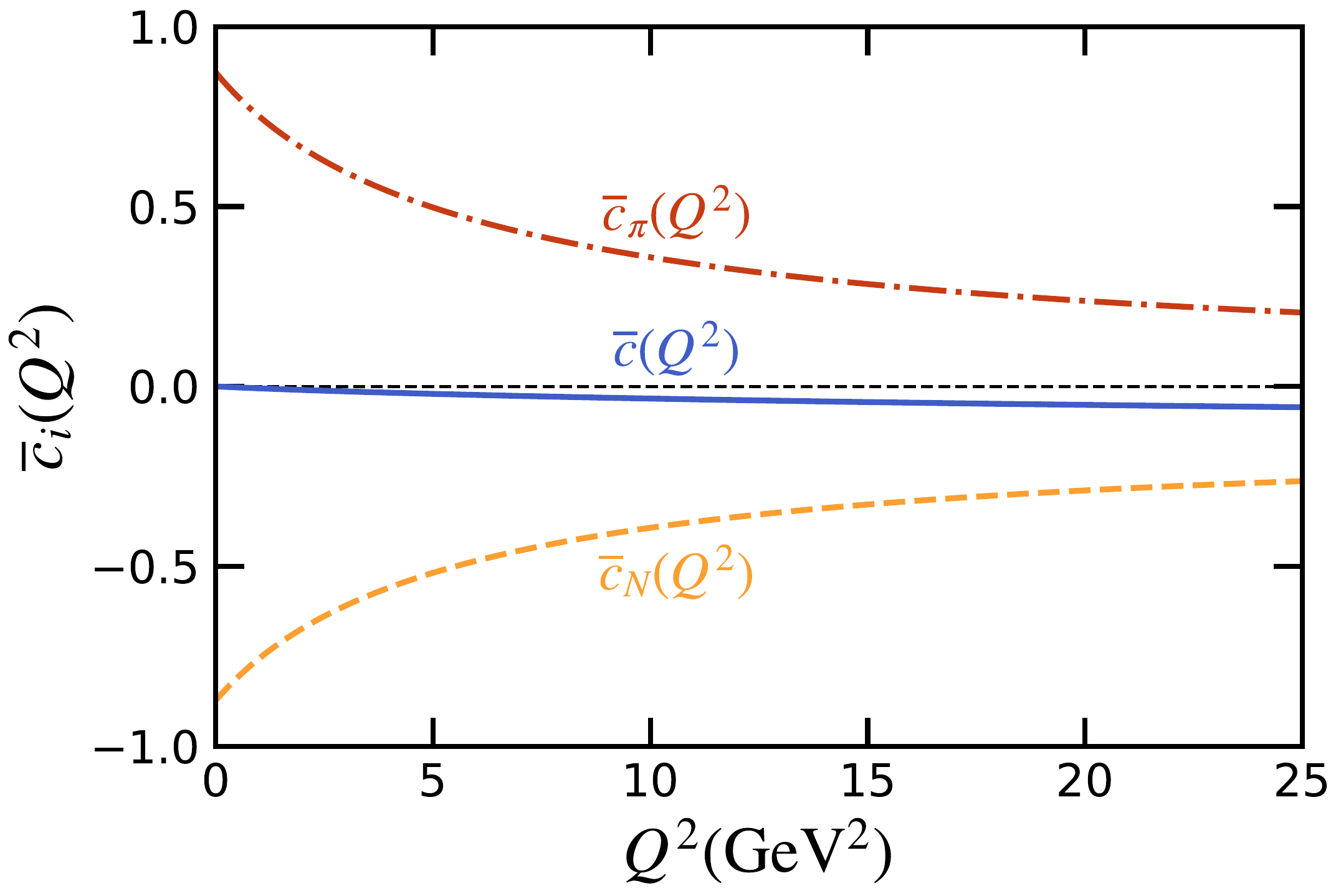}}
\caption{The same as Fig.~\ref{fig:GFFs}, except with the coupling $\alpha=2.0$. }
\label{fig:GFFs_alpha2p0}
\end{figure}

Figure~\ref{fig:GFFs_LCPT} compares the GFFs obtained from leading-order LCPT and those from our non-perturbative solution with a Fock sector truncation up to three-body (one nucleon plus two pions) at $\alpha = 0.5$, where $\alpha$ is the dimensionless Yukawa coupling $\alpha = g^2/(16\pi m^2)$ \cite{Li:2015iaw}.  As mentioned, the three-body truncation results are well converged \cite{Li:2015iaw, Duan:2024dhy}. Fig.~\ref{fig:GFFs_ADC_LCPT} compares the total GFFs $A(Q^2)$, $D(Q^2)$ and $\bar c(Q^2)$. The perturbative and non-perturbative results for $A(Q^2)$ and $D(Q^2)$ are in reasonable agreement. In LCPT, $\bar c(Q^2)$ vanishes as expected, whereas in our non-perturbative calculation, $\bar c(Q^2)$ has a small but non-vanishing value. Fig.~\ref{fig:GFF_Di_LCPT} compares the GFF $D_i(Q^2)$ obtained from LCPT and from our non-perturbative solution. Similarly, Fig.~\ref{fig:GFF_ci_LCPT} compares the GFF $\bar c_i(Q^2)$ obtained from LCPT and from our non-perturbative solution. While the perturbative and non-perturbative results for $A_i(Q^2)$ (not shown) and $D_i(Q^2)$ are in reasonable agreement as expected, GFF $\bar c_i(Q^2)$ exhibits a visible difference between perturbative and non-perturbative results at the coupling $\alpha = 0.5$. The reason is that $\bar c_i(Q^2)$ contains a UV divergence, which magnifies the difference between the perturbative and non-perturbative results. 
Figures~\ref{fig:GFFs} and \ref{fig:GFFs_alpha2p0} present the non-perturbative results at two stronger couplings $\alpha = 1.0$ and $\alpha = 2.0$. 

Figure~\ref{fig:GFF_Ai} shows GFFs $A_N(Q^2)$, $A_\pi(Q^2)$ and their sum $A(Q^2)$ at a non-perturbative coupling $\alpha = 1.0$. As is shown in Ref.~\cite{Cao:2023ohj}, at large $Q^2$, both $A(Q^2)$ and $A_N(Q^2)$ approach the one-body limit $Z$, where $Z$ is the one-body probability, indicating a pointlike nucleon core. And $A_\pi(Q^2)$ approaches zero. 
At small $Q^2$, the slope of $A_i(Q^2)$ in the forward limit controls the root-mean-square (rms) matter radius $r_{Ai}^2 = -6A'_i(0)/A_i(0)$.  At $\alpha = 1.0$, the pion matter radius $r_{A,\pi} = 0.51 \,\mathrm{fm}$ is much larger than the nucleon matter radius $r_{A,N} = 0.09 \,\mathrm{fm}$, consistent with the pion cloud picture. These numbers are tabulated in Table~\ref{table:numbers} for further reference. GFFs $A_N(Q^2)$, $A_\pi(Q^2)$ and $A(Q^2)$ at a larger coupling $\alpha = 2.0$ are shown in Figure~\ref{fig:GFF_Ai_alpha2p0}. 

Figure~\ref{fig:GFF_Di} shows GFFs $D_N(Q^2)$, $D_\pi(Q^2)$ and their sum $D(Q^2)$ at a non-perturbative coupling $\alpha = 1.0$. As is shown in Ref.~\cite{Cao:2023ohj}, at large $Q^2$, both $D(Q^2)$ and $D_N(Q^2)$ approach the one-body limit $-Z$, consistent with the existence of a pointlike nucleon core, whereas $D_\pi(Q^2)$ approaches zero. 
At small $Q^2$, the slope of $D_i(Q^2)$ in the forward limit controls the rms mechanical radius $r_{Di}^2 = -6D'_i(0)/D_i(0)$ \cite{Kumano:2017lhr}.  At $\alpha = 1.0$, the pion mechanical radius $r_{D,\pi} = 1.1 \,\mathrm{fm}$ is much larger than the nucleon mechanical radius $r_{D,N} = 0.15 \,\mathrm{fm}$, indicating a pion cloud outside. Note that the mechanical radius is much larger than the matter radius for both the nucleon and the pion, as shown by Table~\ref{table:numbers}. 
Similar results for $\alpha = 2.0$ are shown in Figure~\ref{fig:GFF_Di_alpha2p0}, and listed in Table~\ref{table:numbers}.

Figure~\ref{fig:GFF_ci} shows GFFs $\bar c_N(Q^2)$, $\bar c_\pi(Q^2)$ and their sum $\bar c(Q^2) = \sum_{i=N,\pi} \bar c_i(Q^2)$ extracted from $T^{12}_i$ and $T^{+-}_i$ at a non-perturbative coupling $\alpha = 1.0$.  Similar results for $\alpha = 2.0$ are shown in Figure~\ref{fig:GFF_ci_alpha2p0}. 
Note that $\bar c_{N,\pi}$ is renormalization scheme and scale dependent, and we have adopted the Pauli-Villars regularization with a Pauli-Villars mass $\mu_\textsc{pv} = 15\,\mathrm{GeV}$ as mentioned. From GFFs $A_i, D_i$ and $\bar c_i$, we can obtain the individual energy density as \cite{Lorce:2017xzd},
\begin{align}
U_i =\,& M\int \frac{\dd^2q_\perp}{(2\pi)^2} e^{i\vec q_\perp \cdot \vec r_\perp}
\Big\{ A_i(q^2_\perp) + \bar c_i(q_\perp^2)  \\
& 
+ \frac{q^2_\perp}{4M^2} \Big[A_i(q^2_\perp) + D_i(q^2_\perp)\Big] 
 \Big\}, \nonumber \\
 \equiv\,& e_i + \Lambda_i,
\end{align}
where, $e_i$ is the internal energy density and $\Lambda_i$ is the external stress. 
The total individual energy, i.e. individual energy density integrated over the transverse space, is given by the GFFs in the forward limit. 
From the values listed in Table~\ref{table:numbers}, we obtain a mass decomposition for our strongly coupled nucleon: the nucleon d.o.f. contributes to 52\% of the total mass while the contribution from the pion d.o.f. is 48\% at $\mu_\textsc{pv} = 15\,\mathrm{GeV}$ and $\alpha = 1.0$. 
If we only consider the internal energies $e_i$, the bare nucleon contributes 90\% and the pion cloud contributes 10\%. 
As the coupling $\alpha$ increases, both $A_\pi(0)$ and $\bar c_\pi(0)$ increase and the fractional mass of the pion d.o.f. increases, and even goes beyond unity. On the other hand, the fractional internal energies remain positive and below unity. 

\begin{table}
    \caption{Gravitational form factors in the forward limit $Q^2 = 0$ as well as the related radii $r_{Ai}^2=-6A'_i(0)/A_i(0)$ and $r_{Di}^2 = -6D'_i(0)/D_i(0)$ at selected couplings $\alpha = 0.5, 1.0, 2.0$. 
    For $\alpha = 0.5$, we also attach light-cone perturbative theory (LCPT) results for comparison. }
    \label{table:numbers}
    \begin{tabular}{p{1.5cm}p{1.5cm} m{1.5cm}p{1.5cm}p{1.5cm}}
    \toprule
     & & $\mathrm{nucleon}$ & $\mathrm{pion}$ & $\mathrm{total}$ \\
     \hline
     \multirow{5}{*}{\makecell{$\alpha = 0.5$ \\ (LCPT)}} & $r_A$ (fm) & $0.076$    & $0.49$ & $0.16$\\
    &  $r_D$ (fm) & $0.11$      & $1.1$ & $0.82$\\
    &  $A_i(0)$ & $0.92$      & $0.080$ & $1.0$ \\
    &  $D_i(0)$ & $-1.2$      & $-1.4$ & $-2.6$ \\
    &  $\bar{c}_i(0)$ & $-0.29$      & $0.29$ & $0.0$ \\
    \hline
    \multirow{5}{*}{$\alpha = 0.5$} & $r_A$ (fm) & $0.061$ & $0.50$ & $0.13$\\
    & $r_D$ (fm) & $0.11$ & $1.1$ & $0.75$ \\
    & $A_i(0)$ & $0.95$ & $0.050$ & $1.0$ \\
    & $D_i(0)$ & $-1.1$ & $-0.97$ & $-2.1$ \\
    & $\bar{c}_i(0)$ & $-0.19$ & $0.19$ & $0.0$ \\
    \hline
    \multirow{5}{*}{$\alpha = 1.0$} & $r_A$ (fm) & $0.086$    & $0.51$ & $0.18$\\
    &  $r_D$ (fm) & $0.15$      & $1.1$ & $0.86$\\
    &  $A_i(0)$ & $0.90$      & $0.10$ & $1.0$ \\
    &  $D_i(0)$ & $-1.3$      & $-2.0$ & $-3.2$ \\
    &  $\bar{c}_i(0)$ & $-0.38$      & $0.38$ & $0.0$ \\
     \hline
     \multirow{5}{*}{$\alpha = 2.0$}  &  $r_A$ (fm) & $0.12$    & $0.53$ & $0.26$\\
      &   $r_D$ (fm) & $0.21$      & $1.1$ & $0.93$\\
     &    $A_i(0)$ & $0.81$      & $0.19$ & $1.0$ \\
      &   $D_i(0)$ & $-1.6$      & $-4.1$ & $-5.8$ \\
      &   $\bar{c}_i(0)$ & $-0.87$      & $0.87$ & $0.0$ \\
    \botrule
    \end{tabular}
\end{table}

The sum of GFF $\bar c_i$, $\bar c(Q^2) = \sum_i \bar c_i(Q^2)$, is of special interest. In the continuum limit, current conservation $\partial_\mu T^{\mu\nu} = 0$ requires that $\bar c(Q^2)$ vanishes for all $Q^2$, i.e. there is no external force for a self-bound system. However, in our result, $\bar c(Q^2)$ is non-vanishing within the numerical precision, as shown in Figs.~\ref{fig:GFF_ci} and \ref{fig:GFF_ci_alpha2p0}. Fortunately, the value of $\bar c(Q^2)$ is small (few percent level) and scale independent, indicating that it does not contain uncanceled divergences. Furthermore, it vanishes in the forward limit $Q^2 = 0$, i.e. $\bar c(0) = 0$, indicating that the net external force is zero, consistent with our previous analysis in Ref.~\cite{Cao:2024rul}. 

The non-vanishing $\bar c(Q^2)$ represents an isotropic spurious external force distribution on the transverse plane. Note that because $\bar c(0)=0$, the light-front energy is still conserved after integrating the light front energy density $T^{+-}$ over the entire space. For the scalar Yukawa model, the force is attractive at the center $r_\perp = 0$ and repulsive elsewhere -- the pion cloud is slightly stretched outward. 
Since this spurious force originates from the Fock space truncation, which implicitly breaks the Poincaré symmetry, we expect that incorporating higher Fock sector contributions would reduce its magnitude \cite{Zhang:2025wli}, which is already very small in the present work.

\section{Summary}\label{sect:summary}

In this work, we investigate the gravitational form factors of a strongly coupled scalar nucleon as described by the 3+1 dimensional scalar Yukawa theory. As a continuation of our previous work \cite{Cao:2023ohj}, we extract the individual gravitational form factors $A_i(Q^2)$, $D_i(Q^2)$ and $\bar c_i(Q^2)$ for the nucleon and the pion degrees of freedom, following the covariant analysis of the hadronic matrix element on the light front \cite{Cao:2024rul}. These observables allow us to establish a coupled multi-fluid picture of the system, in which we can decompose the quantum numbers of the physical particle into the nucleon and the pion degrees of freedom. For example,  the nucleon d.o.f. contributes to 52\% of the physical scalar nucleon mass while the pion d.o.f. contributes to the rest 48\%, at the coupling $\alpha = 1.0$ and a UV resolution $\mu_\textsc{pv} = 15\,\mathrm{GeV}$. 

Our work demonstrates a systematic approach to dissect a strongly coupled quantum system in (3+1) dimensions from the underlying light-front wave functions. This approach provides not only the numbers that can be tested in experiments whenever available, but also a transparent microscopic picture of the system. 
While the many-body wave functions we employed in this work are relatively simple, high-quality light-front wave functions for QCD bound states are being made available with the advance of both theory and computational techniques \cite{Hornbostel:1988fb, Vary:2009gt, Honkanen:2010rc, Chabysheva:2011ed, Ji:2012ux, Chang:2013pq, Zhao:2014xaa, Wiecki:2014ola, Li:2015zda, Lamm:2016djr, Hiller:2016itl, Li:2017mlw, Tang:2018myz, Tang:2019gvn, Jia:2018ary, Qian:2020utg, Kreshchuk:2020aiq, Kreshchuk:2020kcz, Xu:2021wwj, Lan:2021wok, Ji:2021znw, Kuang:2022vdy, Qian:2021jxp}. Methods we developed in our recent works will be useful to gain insights into hadronic structures from QCD. 

\section*{Acknowledgements}

The authors acknowledge fruitful discussions with S. Xu, C. Mondal, S. Nair, X. Zhao and V.A. Karmanov.
Y.L. is supported by the National Natural Science Foundation of China (NSFC) under Grant No.~12375081, by the Chinese Academy of Sciences under Grant No.~YSBR-101. 


\end{document}